\documentclass{IEEEtaes}

\usepackage{color,array,amsthm}
\usepackage{graphicx}

\jvol{XX}
\jnum{XX}
\jmonth{XXXXX}
\paper{1234567}
\pubyear{2026}
\doiinfo{TAES.2026.Doi Number}

\usepackage[]{graphicx}    
\newcommand{\ignore}[1]{}  

\usepackage{amsmath,amssymb,amsfonts}
\usepackage{bm}
\usepackage[ruled,vlined,linesnumbered]{algorithm2e} 
\SetKwComment{tcp}{$\triangleright$ }{}              

\usepackage{float}
\usepackage[labelformat=simple]{subcaption}

\usepackage{mathtools}
\usepackage{siunitx}
\usepackage{multirow}
\usepackage{wasysym}
\usepackage{csquotes}
\usepackage{caption}

\usepackage{mathtools, cuted}
\usepackage{lipsum, color}

\let\labelindent\relax   
\usepackage{enumitem}
\setlist[enumerate]{leftmargin=0.8em,labelsep=0.5em,itemsep=0pt,topsep=2pt,parsep=0pt}

\usepackage{amsthm}
\usepackage[english]{babel}

\newcolumntype{I}{!{\vrule width 1.5pt}}

\newcommand{\newtext}[1]{\textcolor{black}{#1}}
\newcommand{\newcaption}[1]{\caption{\textcolor{black}{#1}}}

\newenvironment{editsection}[1]
  {\begingroup\color{black}}
  {\endgroup}

\begin{document}

\title{Ephemeris and Almanac Design for Lunar Navigation Satellites} 

\author{KEIDAI IIYAMA}
\affil{Stanford University, Stanford, CA, USA} 

\author{GRACE GAO}
\affil{Stanford University, Stanford, CA, USA}


\receiveddate{Manuscript received XXXXX 00, 0000; revised XXXXX 00, 0000; accepted XXXXX 00, 0000.\\
This work was supported in part by the Nakajima Foundation }

\corresp{{\itshape (Corresponding author: Grace Gao)}. }

\authoraddress{The authors are with Stanford University, Stanford, CA 94305 USA 
(e-mail: \href{kiiyama@stanford.edu}{kiiyama@stanford.edu}; \href{gracegao@stanford.edu}{gracegao@stanford.edu}).}


\markboth{IIYAMA AND GAO}{EPHEMERIS AND ALMANAC FOR LUNAR NAVIGATION SATELLITES}
\maketitle

\begin{abstract}
This paper presents almanac and ephemeris message representation for lunar navigation satellites supporting the Lunar Augmented Navigation System (LANS). The proposed method combines osculating orbital elements, Chebyshev polynomials, and Fourier series to efficiently represent lunar satellite trajectories subject to complex perturbations from lunar gravity and third-body effects. For the ephemeris, a hybrid Chebyshev–Fourier formulation improves fitting accuracy over long arcs while maintaining message compactness under the datasize constraint of the LunaNet Interoperability Specification. For the almanac, a compact low-order polynomial and Fourier model is introduced to capture mid-term orbital variations over a 15-day fitting arc. The approach is validated for multiple orbit regimes, including 30-hour, 24-hour, and 12-hour elliptical Lunar frozen orbits (ELFOs) and a 6-hour polar orbit. Results show that the proposed framework achieves sub-meter position and sub-millimeter-per-second velocity fitting errors within the 900-bit limit for 6-hour ephemeris arcs, and almanac fitting accuracy sufficient for reliable satellite visibility identification in warm-start operations. 
\end{abstract}

\begin{IEEEkeywords}
{Lunar positioning, navigation, and timing (PNT), almanac, ephemeris, GNSS}
\end{IEEEkeywords}

\section{Introduction}
\label{sec:introduction}

The pace of lunar exploration has accelerated rapidly in recent years, with more than 40 missions planned by space agencies and private entities between 2024 and 2030~\cite{PlanetarySocietyMoon2024}. 
To support these activities, robust and precise positioning, navigation, and timing (PNT) services are indispensable for orbit determination, surface mobility, logistics, and communication across the lunar environment. 
In response, space agencies are jointly developing LunaNet---an interoperable ``network of networks'' designed to provide navigation and communication services both in lunar orbit and on the surface~\cite{Israel2020}. 
A key component of LunaNet’s navigation services is the Lunar Augmented Navigation Service (LANS), a constellation of lunar-orbiting navigation satellites that will broadcast Augmented Forward Signals (AFS) to enable users to determine their position and time~\cite{LNISv5}.
National Aeronautics and Space Administration (NASA), European Space Agency (ESA), and Japan Aerospace Exploration Agency (JAXA) each plan to deploy their own LANS constellations as part of LunaNet: the Lunar Communications Relay and Navigation Systems (LCRNS)~\cite{gramling2024}, the Lunar Communications and Navigation Service (LCNS)~\cite{traveset2024}, and the Lunar Navigation Satellite System (LNSS)~\cite{Murata2022}, respectively.

Analogous to terrestrial Global Navigation Satellite Systems (GNSS), LANS satellites will broadcast navigation messages containing almanac and ephemeris parameters that allow users to compute satellite position, velocity, and clock offset at any given epoch.
However, designing almanac and ephemeris messages for lunar navigation satellites introduces challenges fundamentally different from those of terrestrial GNSS. 
LunaNet satellites operate in highly elliptical orbits, such as elliptical lunar frozen orbits (ELFOs), which experience distinct dynamical regimes within each revolution: strong third-body perturbations near apolune and significant gravitational anomalies near perilune~\cite{Cortinovis2024Ephemeris}. 
As a result, the satellite motion cannot be accurately represented using simple Keplerian elements with small harmonic corrections, as commonly done for terrestrial systems~\cite{GPS_book}. 
Moreover, the limited number of tracking assets restricts the frequency of orbit updates, making it desirable for each ephemeris segment to remain valid over longer durations than in terrestrial GNSS.
At the same time, the navigation message budget is highly constrained: according to the LunaNet Interoperability Specification, only about 900~bits are available for the ephemeris portion of each broadcast frame after allocating bits for clock, health, and integrity information~\cite{DeOliveiraSalgueiro2025LCNS}. 
Meanwhile, the position and velocity accuracy requirements are stringent, with thresholds of 40~m and 10~mm/s (95th percentile), respectively, including orbit determination and clock-drift errors~\cite{LNISv5}. 
LCRNS imposes even tighter requirements, specifying a Signal-in-Space Error (SISE) of 13.43~m (3$\sigma$) in position and 1.2~mm/s (3$\sigma$, 10~s) in velocity~\cite{LunarSRD2022}. 
Thus, the ephemeris representation must balance accuracy, validity duration, and message compactness.

\begin{table*}[htb]
\caption{\newtext{Summary of Prior Works on Ephemeris Design for Lunar Navigation Satellites (OE: Osculating Elements, EA: Empirical Acceleration, Cheby: Chebyshev)}}
\label{tab:lunar_ephemeris}
\centering
\begin{tabular}{l|l|l|l|c|c|c}
\hline
\textbf{Reference} & \textbf{Ephemeris} & \textbf{Almanac} & \textbf{Orbits} & \textbf{Position} & \textbf{Velocity} & \textbf{Datasize} \\
\hline
\cite{Cortinovis2024Ephemeris} & Cheby & -- & ELFO (12~hr), LLO & \checkmark & \checkmark & \checkmark \\
\cite{Bury2025LCNS} & OE + EA, Cheby & -- & ELFO (23.95~hr) & \checkmark & \checkmark & \checkmark \\
\cite{Hartigan2025LNSS} & OE + Cheby & -- & ELFO (24.37~hr) & \checkmark & -- & -- \\
\cite{DeOliveiraSalgueiro2025LCNS} & OE + Cheby & -- & ELFO (12, 18, 24~hr), DRO & \checkmark & -- & \checkmark \\
This work & OE + Cheby + Fourier & OE (Polynomial + Fourier) & ELFO (12, 24, 30~hr), Polar & \checkmark & \checkmark & \checkmark \\
\hline
\end{tabular}%
\end{table*}

Several studies have explored approaches to parameterize lunar navigation satellite orbits.
\cite{Cortinovis2024Ephemeris} demonstrated a Chebyshev-based state-fitting method that directly fits the Cartesian position and velocity components subject to state constraints, achieving meter-level accuracy for a 12-hour ELFO. 
Similarly, \cite{Bury2025LCNS} proposed using Chebyshev polynomials and osculating orbital elements augmented with empirical accelerations to model the orbits of LCNS satellites in 23.95-hour ELFOs. 
\cite{Hartigan2025LNSS} and \cite{DeOliveiraSalgueiro2025LCNS} introduced hybrid ephemeris representations combining osculating elements and Chebyshev polynomials, demonstrating that such formulations can achieve longer fitting durations and smaller data sizes than Chebyshev-only methods. 
These prior studies are summarized in Table~\ref{tab:lunar_ephemeris}.

While these prior studies have established valuable foundations for lunar ephemeris design, their achievable validity durations remain limited—typically two hours for 12-hour ELFOs and four hours for 24-hour ELFOs~\cite{DeOliveiraSalgueiro2025LCNS}. 
Given the scarcity of lunar tracking assets, extending the ephemeris validity (e.g., to four hours or longer) would be beneficial to reduce the operational burden of frequent updates. 
Furthermore, none of the previous works have examined almanac design for lunar navigation satellites, which is essential for cold- and warm-start receiver operations.

This paper introduces a unified framework for ephemeris and almanac design for LunaNet satellites. 
The proposed ephemeris representation integrates osculating orbital elements with Chebyshev polynomials and Fourier series to capture both short- and mid-term orbital perturbations, improving fitting accuracy over longer intervals compared to prior methods. 
For the almanac, a compact polynomial–Fourier model is developed to represent the long-term evolution of osculating elements and argument of latitude, achieving sub-degree median latitude and longitude errors over 15-day arcs—sufficient for warm-start satellite visibility identification. 
We validate the ephemeris and almanac generation across multiple orbit regimes: 12-, 24-, and 30-hour ELFOs and low-altitude polar orbits.

The remainder of this paper is organized as follows. 
Section~\ref{sec:orbit_analysis} describes the reference frames, dynamics models, and target lunar orbits used in the analysis. 
Section~\ref{sec:ephemeris_almanac_representation} presents the proposed ephemeris and almanac representations. 
Section~\ref{sec:optimization} details the fitting methodology and message-size computation. 
Sections~\ref{sec:results_ephemeris} and~\ref{sec:results_almanac} present the ephemeris and almanac fitting results, respectively. 
Finally, Section~\ref{sec:conclusion} concludes with key findings and directions for future work.
\section{Orbit Analysis}
\label{sec:orbit_analysis}

\subsection{Coordinate Frames and Dynamics}
\label{sec:frame_conversions}
We consider three coordinate frames for the orbit analysis: the Moon-Centered Inertial (MCI) frame, the Moon Principal Axis (PA) frame, and the Principal Axis Inertial (PAI) frame.

The MCI frame is defined with its origin at the Moon’s center of mass, and its axes are aligned with those of the Geocentric Celestial Reference Frame (GCRF) at a reference epoch. This frame serves as the inertial reference for orbit propagation and dynamical modeling.

The PA frame is a rotating frame whose axes are aligned with the Moon’s principal axes of inertia at each epoch $t$. This frame rotates with the Moon and is used as the output frame for the satellite position and velocity in the ephemeris representation~\cite{Folta2022Astrodynamics}.

The PAI frame is also centered at the Moon’s center of mass, but its axes are aligned with the instantaneous orientation of the Moon’s principal axes of inertia at each time $t$, without accounting for the rotational motion of the Moon itself. Because the Moon’s principal axes slowly precess over time, a distinct PAI frame exists at each epoch. This quasi-inertial frame is used during the fitting process to generate osculating orbital elements.

The position vectors in the MCI, PAI, and PA frames are denoted as $\mathbf{r}^{\mathrm{MCI}}$, $\mathbf{r}^{\mathrm{PAI}}$, and $\mathbf{r}^{\mathrm{PA}}$, respectively. Similarly, the velocity vectors are denoted as $\mathbf{v}^{\mathrm{MCI}}$, $\mathbf{v}^{\mathrm{PAI}}$, and $\mathbf{v}^{\mathrm{PA}}$. The positions in the PA and PAI frames are identical, but their velocities differ by the cross product of the angular velocity of the PA frame with respect to the inertial frame and the position vector. The transformations between the MCI, PAI, and PA frames are summarized in~\cite{Folta2022Astrodynamics}.

The lunar orbit propagation is performed in the MCI frame using a dynamics model that includes the Moon’s gravitational field up to degree and order 80, as well as third-body perturbations from the Earth and the Sun~\cite{sat_dynamics}. 

\subsection{\newtext{Target Orbits}}
\label{sec:target_orbits}

\begin{editsection}

Four distinct lunar orbit regimes are considered in this study: a 30-hour Elliptical Lunar Frozen Orbit (ELFO), a 24-hour ELFO, a 12-hour ELFO, and a 6-hour near-circular polar orbit. 
The ELFO configurations correspond to the reference and target orbits for the Lunar Communication Relay and Navigation System (LCRNS)~\cite{NASA2025LCRNSRefConst3_1}, the Lunar Communication and Navigation System (LCNS)~\cite{Melman2025LANSInteroperability}, and the Lunar Navigation Satellite System (LNSS)~\cite{Murata2022}, respectively. 
The polar orbit represents a notional design for the LANS interoperability demonstration mission~\cite{Melman2025LANSInteroperability}.

The initial orbital elements for these orbits are summarized in Table~\ref{tab:initial_conditions}. 
These orbits span a representative range of lunar navigation geometries, covering both elliptical frozen and low-altitude polar regimes, and serve as test cases for the proposed ephemeris and almanac design framework.

\begin{table*}[htb]
\newcaption{Initial Conditions for Target Lunar Orbits ($a$: semi-major axis, $e$: eccentricity, $i$: inclination, $\Omega$: longitude of ascending node, $\omega$: argument of periapsis, $\nu$: true anomaly)}
\label{tab:initial_conditions}
\centering
\resizebox{\textwidth}{!}{%
\begin{tabular}{l|c|l|c|c|c|c|c|c|c}
\hline
\textbf{Orbit} & \textbf{Period} & \textbf{Initial Epoch} & \textbf{Frame} & $\boldsymbol{a}$ (km) & $\boldsymbol{e}$ & $\boldsymbol{i}$ (deg) & $\boldsymbol{\Omega}$ (deg) & $\boldsymbol{\omega}$ (deg) & $\boldsymbol{\nu}$ (deg) \\
\hline
LCRNS (ELFO) & 30 hr & 2027-03-01 00:00:00 UTC & PAI & 11315.94 & 0.692 & 59.373 & 321.019 & 92.494 & 0.000 \\
LCNS (ELFO) & 24 hr & 2027-01-01 00:00:00 TDB & PAI & 9748.14 & 0.700 & 50.638 & 91.603 & 94.344 & 2.344 \\
LNSS (ELFO) & 12 hr & 2027-01-01 00:00:00 TDB & PAI & 6541.40 & 0.600 & 62.940 & 304.170 & 90.013 & 0.000 \\
Polar & 6 hr & 2027-01-01 00:00:00 TDB & PAI & 3870.00 & 0.000 & 89.986 & 34.191 & 67.166 & 7.026 \\
\hline
\end{tabular}%
}
\end{table*}
\end{editsection}


\begin{figure*}[ht!]
    \centering
    \includegraphics[width=\textwidth]{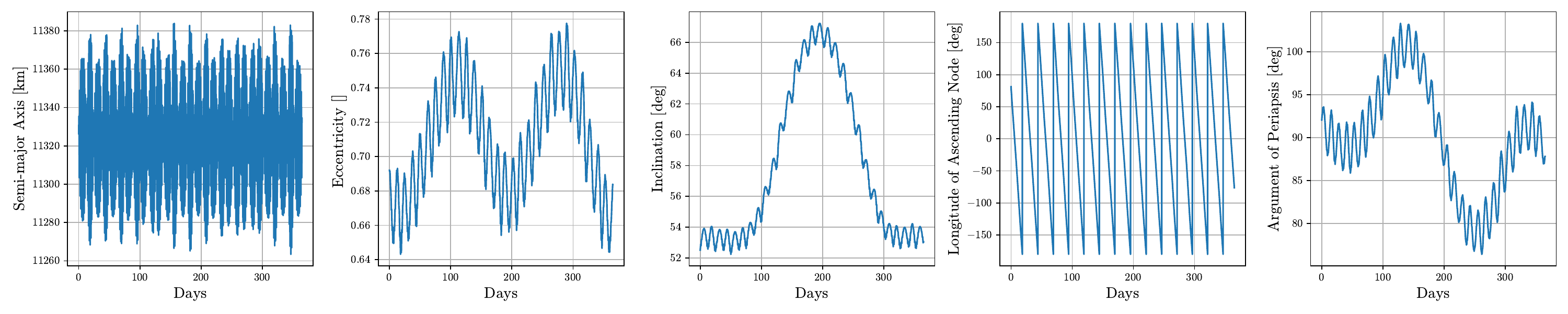}
    \caption{Time histories of osculating orbital elements for the 30-hour ELFO (LCRNS) over 365 days.}
    \label{fig:elfo_orbital_elements}
\end{figure*}

\begin{figure*}[ht!]
    \centering
    \includegraphics[width=\textwidth]{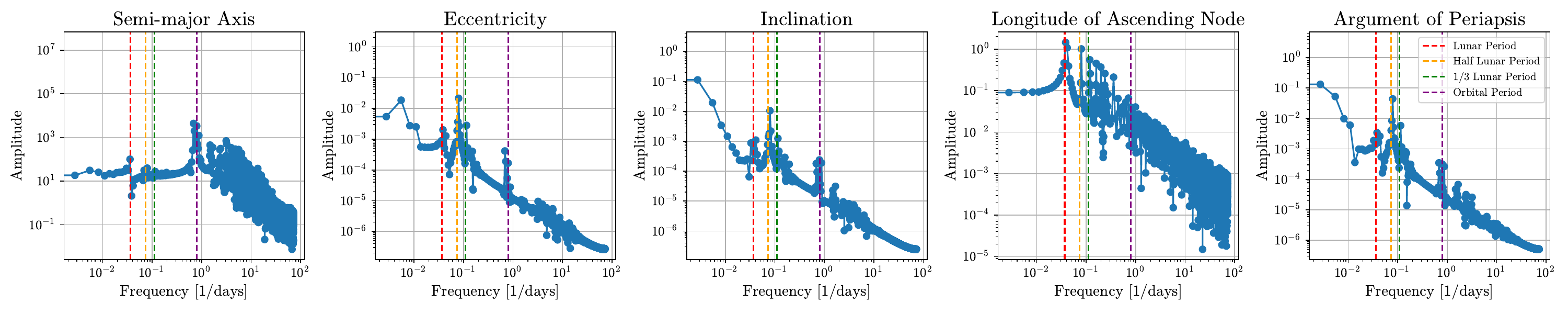}
    \caption{Fourier spectra of the osculating orbital elements for the 30-hour ELFO (LCRNS) over 365 days.}
    \label{fig:fft_elfo_orbital_elements}
\end{figure*}

\begin{figure*}[ht!]
    \centering
    \includegraphics[width=\textwidth]{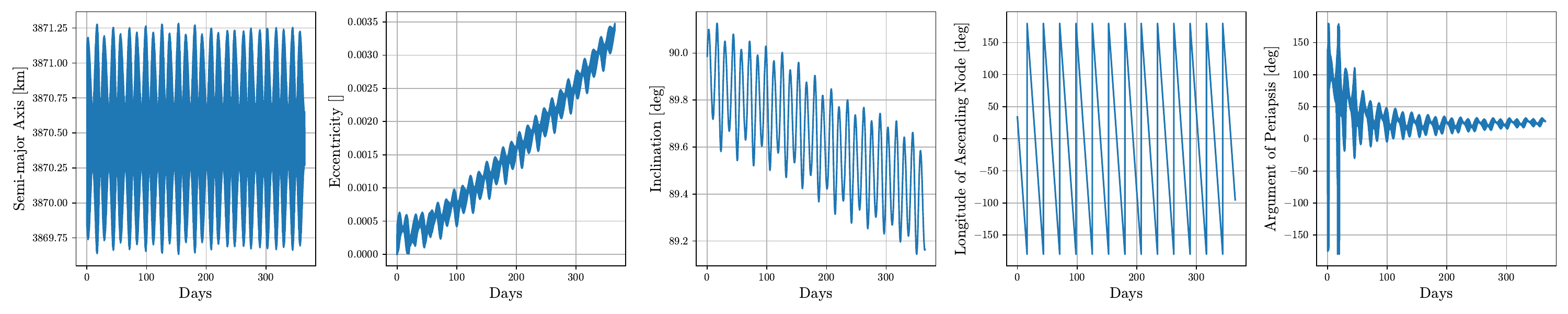}
    \caption{Time histories of osculating orbital elements for the 6-hour polar orbit over 365 days.}
    \label{fig:polar_orbital_elements}
\end{figure*}

\begin{figure*}[ht!]
    \centering
    \includegraphics[width=\textwidth]{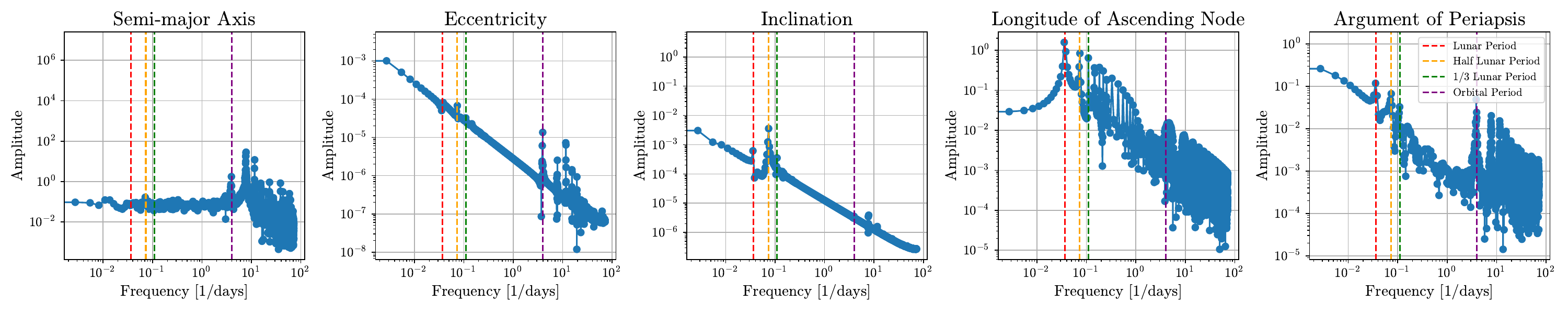}
    \caption{Fourier spectra of the osculating orbital elements for the 6-hour polar orbit over 365 days.}
    \label{fig:fft_polar_orbital_elements}
\end{figure*}

\subsection{Orbit Propagation Analysis}
\label{sec:orbit_propagation_analysis}





Fig.~\ref{fig:elfo_orbital_elements} presents the time histories of the osculating orbital elements (excluding mean anomaly) for the 30-hour Elliptical Lunar Frozen Orbit (ELFO) corresponding to the LCRNS reference orbit, propagated over 365 days. 
As designed, the semi-major axis, eccentricity, inclination, and argument of periapsis remain nearly constant throughout the year, confirming the long-term stability of the frozen-orbit configuration. 
In contrast, the longitude of the ascending node exhibits a clear secular regression, primarily driven by the Moon’s non-spherical gravity field and its rotational coupling effects.

The corresponding Fourier spectra, shown in Fig.~\ref{fig:fft_elfo_orbital_elements}, reveal dominant periodic components near half the lunar sidereal period (\(\approx 13.7\)~days) in \(e\), \(i\), \(\Omega\), and \(\omega\). 
These components are consistent with third-body perturbations induced by Earth. 
In contrast, the semi-major axis spectrum exhibits high-frequency power near the orbital period (\(\approx 30\)~hours), reflecting short-period variations arising from higher-order terms in the lunar gravity potential.

The results for the low-altitude 6-hour polar orbit are shown in Figs~\ref{fig:polar_orbital_elements} and~\ref{fig:fft_polar_orbital_elements}. 
Similar to the ELFO case, the semi-major axis remains relatively stable, while the eccentricity, inclination, and argument of periapsis exhibit gradual secular variations due to higher-order lunar gravitational harmonics. 
A strong periodic component near half the lunar sidereal period (\(\approx 13.7\)~days) persists in the inclination spectrum, again indicative of third-body coupling. 
However, for \(e\) and \(\omega\), the dominant spectral power shifts toward the orbital-period band (\(\approx 6\)~hours), highlighting the increasing influence of short-period terms in low-altitude regimes.

\section{Ephemeris and Almanac Representations}
\label{sec:ephemeris_almanac_representation}

\subsection{Ephemeris Representation}
The proposed ephemeris representation combines osculating orbital elements with Chebyshev polynomials and Fourier series to model the satellite's position and velocity over a specified time interval. Similar to the ephemeris design presented in \cite{Hartigan2025LNSS,DeOliveiraSalgueiro2025LCNS}, we represent the two-body motion using five osculating orbital elements: semi-major axis \( a \), eccentricity \( e \), inclination \( i \), longitude of ascending node \( \lambda \), and argument of periapsis \( \omega \), while the mean anomaly \( M \) and argument of latitude \( u \) are computed from these elements and time.

The orbital elements are converted to Cartesian position and velocity in the PA frame. Then, the residuals between the actual ephemeris data and the two-body motion in the PA frame are modeled using Chebyshev polynomials and Fourier series to capture both secular trends and periodic variations.

Chebyshev polynomials of the first kind, denoted \(F_n(\tau)\), form an orthogonal basis on the interval \(\tau \in [-1,1]\), where
\begin{equation}
\tau = \frac{t - t_0}{T_{\mathrm{fit}}},
\end{equation}
with \(T_{\mathrm{fit}}\) the fitting interval and \(t_0\) the reference epoch at its midpoint. 
They are defined recursively as
\begin{equation}
\begin{aligned}
& F_0(\tau) = 1, \quad 
F_1(\tau) = \tau \\
& F_{n+1}(\tau) = 2\tau F_n(\tau) - F_{n-1}(\tau), 
\quad n \ge 1.
\end{aligned}
\end{equation}
Chebyshev polynomials minimize the maximum approximation error (the minimax property)~\cite{Fraser1965}, allowing high-accuracy fitting with a relatively small number of coefficients. 
An \(n\)th-order polynomial requires \(n+1\) coefficients to represent the function over the normalized interval.

The second component of the ephemeris representation is the Fourier series, which captures periodic variations in the satellite's motion, which is given by:
\begin{equation}
f(t) = c_0 + \left[ C_{c} \cos\left( 2 u(t) \right) + C_{sn} \sin\left( 2 u(t) \right) \right],
\end{equation}
where \( c_0 \) is the average value of the function over one period, which will be shared with the Chebyshev polynomials, \( u(t) \) is the argument of latitude at time $t$, and \( C_{cn} \) and \( C_{sn} \) are the Fourier coefficients for the cosine and sine terms, respectively. We use the argument of latitude to capture the periodic effects corresponding to the orbital period. The longer periodic effects which has period of the lunar sidereal period can be sufficiently captured by the \newtext{Chebyshev} polynomial terms.

The algorithm to compute the x, y, z position and velocity in the PA frame from the ephemeris parameters is summarized in Algorithm \ref{alg:ephemeris_to_state}.

\begin{algorithm}
\caption{Ephemeris to State Conversion}
\label{alg:ephemeris_to_state}

Compute the mean anomaly at time \( t \):
\[
M(t) = M_0 + n (t - t_0), \quad n = \sqrt{\mu_{\leftmoon} / a_0^3}.
\]

Solve Kepler’s equation for \(E(t)\):
\[
M(t) = E(t) - e sin E(t)
\]

Compute the true anomaly:
\[
\nu(t) = 2\arctan\!\left(
\sqrt{(1+e_0)/(1-e_0)} \tan (E(t)/2)
\right).
\]

Compute the orbital-plane radius, argument of latitude, and position in the plane:
\begin{equation*}
\begin{aligned}
r(t) = a(t)\big(1 - e_0 \cos E(t)\big), 
\quad
u(t) = w_0 + \nu(t). \\
x^p(t) = r(t)\cos u(t), 
\quad
y^p(t) = r(t)\sin u(t).
\end{aligned}
\end{equation*}

Adjust the longitude of the ascending node:
\[
\lambda(t) = \lambda_0 - \omega_b (t - t_0).
\]

Compute the nominal position in the PA frame:
\begin{equation*}
    \begin{aligned}
        \newtext{x^{\mathrm{PA}}_{\mathrm{OE}}} (t) &= x^p(t)\cos \lambda(t) - y^p(t)\cos i_0 \sin \lambda(t) \\
        \newtext{y^{\mathrm{PA}}_{\mathrm{OE}}} (t) &= x^p(t)\sin \lambda(t) + y^p(t)\cos i_0 \cos \lambda(t) \\
        \newtext{z^{\mathrm{PA}}_{\mathrm{OE}}} (t) &= y^p(t)\sin i_0
    \end{aligned}
\end{equation*}

Compute the nominal velocity in the PA frame:
\begin{equation*}
    \begin{aligned}
\newtext{\dot{x}^{\mathrm{PA}}_{\mathrm{OE}}} (t) &= -x^p(t)\sin \lambda(t) \dot{\lambda}(t) \\
&- y^p(t) \cos i_0 \cos \lambda(t) \dot{\lambda}(t) \\
&+ \dot{x}^p(t)\cos \lambda(t) - \dot{y}^p(t)\cos i_0 \sin \lambda(t) \\
\newtext{\dot{y}^{\mathrm{PA}}_{\mathrm{OE}}} (t) &= x^p(t)\cos \lambda(t) \dot{\lambda}(t) \\
&- y^p(t)\cos i_0 \sin \lambda(t) \dot{\lambda}(t) \\
&+ \dot{x}^p(t)\sin \lambda(t) + \dot{y}^p(t)\cos i_0 \cos \lambda(t) \\
\newtext{\dot{z}^{\mathrm{PA}}_{\mathrm{OE}}} (t) &= \dot{y}^p(t)\sin i_0
    \end{aligned}
\end{equation*}
where 
\begin{equation*}
    \begin{aligned}
    \dot{\lambda}(t) &= -\omega_b, \quad \ \dot{r}(t) = a e \sin E(t) \dot{E} \\
    \dot{u}(t) &= \frac{\dot{E} \sqrt{1 - e^2}}{1 - e \cos E(t)}, \quad \dot{E}(t) = \frac{n}{1 - e \cos E(t)} \\
    \dot{x}^p(t) &= \dot{r}(t)\cos u(t) - r(t)\sin u(t) \dot{u}(t) \\
    \dot{y}^p(t) &= \dot{r}(t)\sin u(t) + r(t)\cos u(t) \dot{u}(t),
\end{aligned}
\end{equation*}

Compute the residuals in the PA frame \( (\xi = \{x, y, z\}) \):
\begin{equation*}
    \begin{aligned}
    \Delta {\xi}^{\mathrm{PA}}(t) &= \sum_{k=0}^N C_{F, k}^{\xi} F_k \left(\frac{t - t_0}{T_{fit}} \right) \\
    &+ C_c^{\xi} \cos(u(t)) + C_s^{\xi} \sin(u(t)) \\
    \Delta \dot{\xi}^{\mathrm{PA}}(t) &= \frac{1}{T_{fit}} \sum_{k=0}^N C_{F, k}^{\xi} \dot{F}_k \left(\frac{t - t_0}{T_{fit}} \right) \\
    &- C_c^{\xi}\sin(u(t)) \dot{u}(t) + C_s^{\xi} \cos(u(t)) \dot{u}(t)
\end{aligned}
\end{equation*}

Compute the position and velocity in the PA frame \( (\xi = \{x, y, z\}) \):
\[
\xi^{\mathrm{PA}}(t) = \newtext{\xi^{\mathrm{PA}}_{\mathrm{OE}}} (t) + \Delta \xi^{\mathrm{PA}}(t), \
\dot{\xi}^{\mathrm{PA}}(t) = \newtext{\dot{\xi}^{\mathrm{PA}}_{\mathrm{OE}}} (t) + \Delta \dot{\xi}^{\mathrm{PA}}(t)
\]

\end{algorithm}

\subsection{Almanac Representation}
\label{sec:almanac}

The almanac provides a simplified, long-term representation of the satellite’s orbit that preserves its essential geometric characteristics while minimizing computational complexity and message size. 
Unlike the ephemeris, which fits short-arc dynamics using residuals from a two-body model, the almanac fitting is applied directly to the osculating orbital elements over extended time intervals of several days. 
In this regime, lunar gravitational harmonics and third-body perturbations introduce significant deviations from two-body motion, making direct element fitting more robust and compact.

To capture both secular and periodic variations, each osculating element \(\xi(t)\) is expressed as a combination of a low-order polynomial and a single-term Fourier series:
\begin{equation}
\begin{aligned}
\xi(t) &= \beta_0^{\xi} + \beta_1^{\xi} t 
+ \left[ C_{c}^{\xi}\cos\!\left(\omega t\right)
+ C_{s}^{\xi}\sin\!\left(\omega t\right) \right], \\
\omega &= 
\begin{cases}
\displaystyle \frac{2\pi}{T_{\mathrm{orb}}}, & (\xi = a), \\[6pt]
\displaystyle \frac{4\pi}{T_{\mathrm{sid}}}, & \newtext{(\xi = e, i, \Omega, M, u)},
\end{cases}
\end{aligned}
\end{equation}
where \(\xi(t)\) denotes the osculating orbital element at time \(t\); 
\(\beta_k^{\xi}\) are the polynomial coefficients of order \(k\); 
and \(C_{c}^{\xi}\) and \(C_{s}^{\xi}\) are the Fourier coefficients for the cosine and sine terms, respectively. 
The frequency \(\omega\) is chosen according to the dominant periodicity observed in the Fourier analysis (Section~\ref{sec:orbit_analysis}): 
for the semi-major axis, the orbital period \(T_{\mathrm{orb}}\) captures short-period variations, 
whereas for the remaining elements \newtext{(\(e, i, \Omega, M, u\))}, the lunar sidereal period \(T_{\mathrm{sid}} = 27.321661~\text{days}\) captures longer-period modulations driven by third-body and gravitational coupling effects.

Each element’s parameter set is defined as
\[
\boldsymbol{\alpha}_{\xi} = 
[\beta_0^{\xi}, \beta_1^{\xi}, C_{c}^{\xi}, C_{s}^{\xi}],
\]
for each osculating element and $u$. 
\newtext{Note that the argument of periapsis $\omega$ is not corrected since the almanac parameters of $u = \nu + \omega$ can capture the fitting residuals.}

This compact polynomial–Fourier formulation enables accurate prediction of satellite geometry over multi-day periods while maintaining a lightweight parameterization suitable for broadcast almanac messages. 
The algorithm to compute the Cartesian position and velocity components \((x, y, z, \dot{x}, \dot{y}, \dot{z})\) in the PA frame from the almanac parameters is provided in Algorithm~\ref{alg:almanac_to_state}.


\begin{algorithm}
\caption{Almanac to State Conversion}
\label{alg:almanac_to_state}

Compute the orbital period used for the computation of the semi-major axis:
\[
T_{orb}(t) = 2\pi \sqrt{\frac{\bar{a}(t)^3}{\mu_{\leftmoon}}}, 
\qquad \bar{a}(t) = \beta_0^a + \beta_1^a.
\]

Compute the angular frequencies for the Fourier terms:
\[
\omega_{orb} = \frac{2\pi}{T_{orb}(t)}, 
\qquad \omega_{sid} = \frac{4\pi}{T_{syn}}.
\]

\newtext{Compute the time-varying orbital elements}:
\begin{equation*}
    \begin{aligned}
        a(t) &= \beta_0^a + \beta_1^a t 
        + \big[ C_c^a \cos(\omega_{orb} t) + C_s^a \sin(\omega_{orb} t) \big] \\
        e(t) &= \beta_0^{e} + \beta_1^{e} t 
        + \big[ C_c^{e} \cos(\omega_{sid} t) + C_s^{e} \sin(\omega_{sid} t) \big] \\
        i(t) &= \beta_0^{i} + \beta_1^{i} t 
        + \big[ C_c^{i} \cos(\omega_{sid} t) + C_s^{i} \sin(\omega_{sid} t) \big] \\
        \lambda(t) &= \beta_0^\lambda + \beta_1^\lambda t 
        + \big[ C_c^\lambda \cos(\omega_{sid} t) + C_s^\lambda \sin(\omega_{sid} t) \big] \\
        &\quad - \omega_b t,
    \end{aligned}
\end{equation*}

Compute the nominal mean anomaly using trapezoidal integration:
\[
n(t) = \sqrt{\frac{\mu_{\leftmoon}}{a(t)^3}},
\qquad M_0(t) = \int_0^t n(\tau)\, d\tau.
\]

Correct the mean anomaly using polynomial and sidereal Fourier terms:
\begin{equation*}
    \begin{aligned}
        M(t) &= M_0(t) + \beta_0^M + \beta_1^M t \\
        &\quad + \big[ C_c^M \cos(\omega_{sid} t) + C_s^M \sin(\omega_{sid} t) \big].
    \end{aligned}
\end{equation*}

Solve Kepler’s equation for  \(E(t)\):
\[
M(t) = E(t) - e sin E(t)
\]

Compute the true anomaly:
\[
\nu(t) = 2\arctan\!\left(
\sqrt{\frac{1+e(t)}{1-e(t)}} \tan \frac{E(t)}{2}
\right).
\]

Compute the argument of latitude:
\begin{equation*}
    \begin{aligned}
        u(t) &= \newtext{\nu(t) + \beta_0^u + \beta_1^u t} \\
        &\quad + \big[ C_c^u \cos(\omega_{sid} t) + C_s^u \sin(\omega_{sid} t) \big].
    \end{aligned}
\end{equation*}

Compute the orbital-plane radius and position:
\begin{equation*}
    \begin{aligned}
        r(t) &= a(t)\big(1 - e(t)\cos E(t)\big) \\
        x^p(t) &= r(t)\cos u(t) \\
        y^p(t) &= r(t)\sin u(t) 
    \end{aligned}
\end{equation*}

Compute the position in the PA frame:
\begin{equation*}
    \begin{aligned}
        x^{\mathrm{PA}}(t) &= x^p(t)\cos \lambda(t) - y^p(t)\cos i(t)\sin \lambda(t) \\
        y^{\mathrm{PA}}(t) &= x^p(t)\sin \lambda(t) + y^p(t)\cos i(t)\cos \lambda(t) \\
        z^{\mathrm{PA}}(t) &= y^p(t)\sin i(t)
    \end{aligned}
\end{equation*}

Compute the velocity in the PA frame by finite differencing \( (\xi = \{x, y, z\}) \):
\[
\dot{\xi}^{\mathrm{PA}}(t) =
\frac{\xi^{\mathrm{PA}}(t + \Delta t) - \xi^{\mathrm{PA}}(t)}{\Delta t}.
\]

\end{algorithm}

\section{Optimization}
\label{sec:optimization}

\subsection{Least Squares Fitting}
\label{sec:least_squares}

Both the ephemeris and almanac parameterizations are optimized via a least-squares fitting procedure. 
Given a set of $N_T$ sampled position and velocity states 
\(\{ \mathbf{X}^{\mathrm{PAI}}(t_m) \}_{m=1}^{N_T}\) 
in the Principal-Axis Inertial (PAI) frame over a fitting interval \([t_{\mathrm{init}}, t_f]\), where 
\(\mathbf{X}^{\mathrm{PAI}}(t_m) = [x^{\mathrm{PAI}}(t_m), y^{\mathrm{PAI}}(t_m), z^{\mathrm{PAI}}(t_m)]^{\top}\),
the goal is to determine the optimal parameter set \(\boldsymbol{\alpha}\) that minimizes the residuals between the sampled data and the model prediction.

The sampling epochs are distributed using Chebyshev–Lobatto nodes~\cite{ChebyshevBook},
\begin{equation}
    t_m = \frac{t_f + t_{\mathrm{init}}}{2}
    + \frac{t_f - t_{\mathrm{init}}}{2}
    \cos\!\left(\frac{m\pi}{N_T}\right),
\end{equation}
to reduce Runge’s phenomenon and improve numerical stability during polynomial fitting. 
The fitting interval duration is denoted \(T_{\mathrm{fit}} = t_f - t_{\mathrm{init}}\).

\subsubsection{Ephemeris Fitting}
\label{sec:ephemeris_fitting}

The ephemeris fitting procedure is performed as follows:

\begin{enumerate}
    \item Define the reference epoch \(t_0 = (t_{\mathrm{init}} + t_f)/2\) and initialize the osculating orbital elements 
    \(\{ a_0, e_0, i_0, \Omega_0, \omega_0, M_0 \}\)
    from the PAI state at \(t_0\).

    \item Using Algorithm~\ref{alg:ephemeris_to_state} (lines~1–6), propagate the nominal two-body motion to obtain the predicted state
    \newtext{\(\hat{\mathbf{X}}^{\mathrm{PA}}_{\mathrm{OE}}(t_m)\)}
    and argument of latitude \(u_m\) at each node \(t_m\).

    \item Compute the residuals between the sampled and nominal states:
    \begin{equation}
    \begin{aligned}
        \Delta \mathbf{X}^{\mathrm{PAI}}(t_m)
        &= \mathbf{X}^{\mathrm{PAI}}(t_m)
        - \newtext{\hat{\mathbf{X}}^{\mathrm{PA}}_{\mathrm{OE}}(t_m)} \\
        & \quad (m = 1, \ldots, N_T).
    \end{aligned}
    \end{equation}

    \item Construct the design matrix \(T\) based on the Chebyshev–Fourier basis:
    \begin{equation}
    \begin{aligned}
        &\newtext{\mathbf{T}} = \\
        &\begin{bmatrix}
            F_0(\tau_1)  & \cdots & F_n(\tau_1) & \cos(2u_1) & \sin(2u_1) \\
            F_0(\tau_2) & \cdots & F_n(\tau_2) & \cos(2u_2) & \sin(2u_2) \\
            \vdots & \ddots & \vdots & \vdots & \vdots \\
            F_0(\tau_{N_T}) & \cdots & F_n(\tau_{N_T}) & \cos(2u_{N_T}) & \sin(2u_{N_T})
        \end{bmatrix},
    \end{aligned}
    \end{equation}
    where \(\tau_m = (t_m - t_0) / T_{\mathrm{fit}} \) and \(F_n(\tau)\) denotes the \(n\)th Chebyshev polynomial.

    \item Solve the least-squares problem to estimate the ephemeris parameters:
    \begin{equation}
        \min_{\boldsymbol{\alpha}}
        \sum_{m=1}^{N_T}
        \left\|
        \Delta \mathbf{X}^{\mathrm{PAI}}(t_m)
        - \newtext{\mathbf{T}_m} \boldsymbol{\alpha}
        \right\|_2^2,
    \end{equation}
    where \newtext{\(\mathbf{T}_m\)} denotes the \(m\)th row of \newtext{\(\mathbf{T}\)}.
\end{enumerate}



\subsubsection{Almanac Fitting}
\label{sec:almanac_fitting}

The almanac parameters are obtained through a similar least-squares process applied directly to each osculating orbital element:

\begin{enumerate}
    \item Compute the osculating elements
    \newtext{\(\{ a(t_m), e(t_m), i(t_m), \Omega(t_m), \omega(t_m), M(t_m) \}\)}
    from the PAI states.

    \item Define the residuals for each element as
    \begin{equation}
        \Delta \xi(t_m) =
        \begin{cases}
            \xi(t_m), & \xi \in \{ a, e, i, \omega \},\\[4pt]
            \xi(t_m) + \omega_b t_m, & \xi = \Omega,
        \end{cases}
    \end{equation}
    where \(\omega_b = 2.6617\times10^{-6}~\mathrm{rad/s}\) is the lunar rotation rate.

    \item Compute the mean semi-major axis over the fitting interval:
    \begin{equation}
        \bar{a} = \frac{1}{N_T}\sum_{m=1}^{N_T} a(t_m).
    \end{equation}

    \item Construct the design matrix for the semi-major axis:
    \begin{equation}
    \begin{aligned}
        \newtext{\mathbf{T}_a} &= \\
        &\begin{bmatrix}
            1 & t_1 & \cos(\omega_{\mathrm{orb}} t_1) & \sin(\omega_{\mathrm{orb}} t_1) \\
            1 & t_2 & \cos(\omega_{\mathrm{orb}} t_2) & \sin(\omega_{\mathrm{orb}} t_2) \\
            \vdots & \vdots & \vdots & \vdots \\
            1 & t_{N_T} & \cos(\omega_{\mathrm{orb}} t_{N_T}) & \sin(\omega_{\mathrm{orb}} t_{N_T})
        \end{bmatrix},
    \end{aligned}
    \end{equation}
    where \(\omega_{\mathrm{orb}} = 2\pi/T_{\mathrm{orb}}\) and \(T_{\mathrm{orb}} = 2\pi \sqrt{\bar{a}^3 / \mu_{\leftmoon}}\).

    \item Solve the least-squares problem for the semi-major axis:
    \begin{equation}
        \min_{\boldsymbol{\alpha}_a}
        \sum_{m=1}^{N_T}
        \|\Delta a(t_m) - \newtext{\mathbf{T_{a,m}}} \boldsymbol{\alpha}_a\|_2^2.
    \end{equation}
    where 

    \item Repeat \newtext{Step~5} for each remaining element
    \newtext{\(\xi \in \{e, i, \Omega\}\)}
    using
    \begin{equation}
    \begin{aligned}
        \mathbf{T}_{\mathrm{siderial}} &= \\
        &\begin{bmatrix}
            1 & t_1 & \cos(\omega_{\mathrm{sid}} t_1) & \sin(\omega_{\mathrm{sid}} t_1) \\
            1 & t_2 & \cos(\omega_{\mathrm{sid}} t_2) & \sin(\omega_{\mathrm{sid}} t_2) \\
            \vdots & \vdots & \vdots & \vdots \\
            1 & t_{N_T} & \cos(\omega_{\mathrm{sid}} t_{N_T}) & \sin(\omega_{\mathrm{sid}} t_{N_T})
        \end{bmatrix},
    \end{aligned}
    \end{equation}
    where \(\omega_{\mathrm{sid}} = 4\pi/T_{\mathrm{sid}}\) and \(T_{\mathrm{sid}} = 27.321661~\text{days}\) is the lunar sidereal period.

    \item \newtext{Compute} the \newtext{predicted} mean anomaly \(\hat{M}_0(t_m)\) using Algorithm~\ref{alg:almanac_to_state} (lines~1–4) and compute
    \(\Delta M = M(t_m) - \hat{M}_0(t_m)\).
    Fit \(\boldsymbol{\alpha}_M\) using design matrix \newtext{\(\mathbf{T}_{\mathrm{sidereal}}\)}.

    \item Compute the predicted true anomaly $\hat{\nu}(t_m)$ using fitted eccentricity $\hat{e}(t_m) = (\mathbf{T}_{sideral})_m \boldsymbol{\alpha}_e$ and mean anomaly $\hat{M}(t_m) = M_0 + (\mathbf{T}_{sidereal})_m \boldsymbol{\alpha}_M $

    \item Finally, compute argument of latitude residuals
    \newtext{\(\Delta u (t_m) = \omega(t_m) + \nu(t_m) - \hat{\nu}(t_m) \)},
    and solve for \(\boldsymbol{\alpha}_u\) using \newtext{\(\mathbf{T}_{\mathrm{sidereal}}\)}.
\end{enumerate}

\subsection{Message Size Computation}
\label{sec:message_size}

The message size required to broadcast the ephemeris or almanac can be estimated following the procedure outlined in~\cite{DeOliveiraSalgueiro2025LCNS}, with additional subprocedures introduced here to compute parameter resolutions efficiently using numerical differentiation.

\begin{enumerate}
    \item Resolution Computation:
    For each parameter (e.g., Chebyshev coefficients, Fourier coefficients, or orbital elements), determine the bit resolution required to ensure that the resulting position error remains below a specified tolerance \(\delta_x\).  
    The process consists of three steps:

    \begin{enumerate}
        \item Numerical sensitivity evaluation: 
        Compute the sensitivity of the satellite position to each parameter by evaluating the partial derivative of the position function with respect to that parameter using finite differences:
        \begin{equation}
        \begin{aligned}
            &\left| \frac{\partial f_{\boldsymbol{\alpha} \rightarrow x}(\boldsymbol{\alpha}, t)}{\partial \alpha_i} \right|
            \approx
            \frac{1}{N_t \epsilon}
            \sum_{m=1}^{N_t} \delta f_m (\boldsymbol{\alpha}) \\
             &\delta f_m (\boldsymbol{\alpha}) = f_{\boldsymbol{\alpha} \rightarrow x}(\boldsymbol{\alpha} + \epsilon \mathbf{e}_i, t_m)
            \\
            & \qquad - f_{\boldsymbol{\alpha} \rightarrow x}(\boldsymbol{\alpha} - \epsilon \mathbf{e}_i, t_m)
        \end{aligned}
        \end{equation}
        where \(f_{\boldsymbol{\alpha} \rightarrow x}(\boldsymbol{\alpha}, t)\) maps the parameter set \(\boldsymbol{\alpha}\) to the satellite position at time \(t\); 
        \(t_m\) (\(m=1,\ldots,N_t\)) are the sampling times; 
        \(\mathbf{e}_i\) is the unit vector corresponding to parameter \(\alpha_i\); 
        and \(\epsilon = 10^{-3}\) is a small perturbation.

        \item Initial bit-depth estimate: 
        The initial estimate of the bit depth \(k_i\) for parameter \(\alpha_i\) is computed as
        \begin{equation}
            k_i = - \left\lceil \log_2 \left( 
            \frac{\delta_x}{
            \left| \frac{\partial f_{\boldsymbol{\alpha} \rightarrow x}}{\partial \alpha_i} \right|}
            \right) \right\rceil,
            \label{eq:k_initial_guess}
        \end{equation}
        where \(\lceil \cdot \rceil\) denotes the ceiling function and \(\delta_x = 0.01~\mathrm{m}\) is the target position accuracy.

        \item Iterative refinement:  
        Starting from the initial estimate in Eq.~\eqref{eq:k_initial_guess}, iteratively adjust \(k_i\) by \(\pm 1\) until the smallest value satisfying the position accuracy constraint is found:
        \begin{equation}
        \begin{aligned}
            k_i &= \arg \min_{k_i}
            \bigg\{
            \max_{s \in \{-1, 1\}}
            \max_{m \in [1, N_t]} \\
            & \| f_{\boldsymbol{\alpha} \rightarrow x}(\boldsymbol{\alpha} + s\,2^{-k_i}\mathbf{e}_i, t_m) - \bar{\mathbf{x}}_m \|_2
            < \delta_x
            \bigg\}
        \end{aligned}
        \end{equation}
        where \(\bar{\mathbf{x}}_m\) is the reference position at time \(t_m\).
    \end{enumerate}

    \item Bit Allocation:  
    Once the required resolution is determined, the number of bits \(b_i\) required to encode each parameter is given by
    \begin{equation}
    \begin{aligned}
        b_i &= \max \left\{
        \left\lceil
        \log_2 \!\left( \alpha_{i,\mathrm{max}} - \alpha_{i,\mathrm{min}} \right)
        + k_i
        \right\rceil, 1
        \right\} \\
        &\quad + b_s + b_m,
    \end{aligned}
    \end{equation}
    where \(\alpha_{i,\mathrm{max}}\) and \(\alpha_{i,\mathrm{min}}\) denote the maximum and minimum values of \(\alpha_i\) across different fitting arcs (with varying starting mean anomalies).  
    The terms \(b_s\) and \(b_m\) represent the bits allocated for sign (1 for signed, 0 for unsigned) and for margin to account for quantization and potential future adjustments, respectively.  
    No margin bits are allocated for the bounded elements \(e, i, \Omega, \omega,\) and \(M\), as their ranges are limited to \([0,1]\) or \([0,2\pi]\).

    \item Total Message Size: 
    The total message size (in bits) is the sum of the individual parameter allocations:
    \begin{equation}
        \text{Message Size}
        = \sum_{i=1}^{N_c} b_i,
    \end{equation}
    where \(N_c\) is the total number of fitted parameters.
\end{enumerate}

\section{Ephemeris Simulation Results}
\label{sec:results_ephemeris}

\subsection{Simulation Setup}
The ephemeris fitting experiments are conducted for four lunar orbit regimes introduced in Section \ref{sec:orbit_analysis}.
We evaluate fitting intervals of 60, 120, 240, 360, and 480~min, and compare three ephemeris parameterizations: \emph{(i)} Chebyshev polynomials only, \emph{(ii)} Chebyshev + osculating elements, and \emph{(iii)} Chebyshev + osculating elements + Fourier series.

For each orbit and fitting interval, we draw 30 starting mean anomalies uniformly over \([0,2\pi)\). Using these 30 fits, we compute the required data size per representation via the bit-allocation method in Section~\ref{sec:message_size}. 
Fitting accuracy is reported as the 95th-percentile position (norm) and velocity (norm) errors evaluated over all points in fitting arcs with different starting mean anomalies.

Sampling for fitting uses one point per minute with Chebyshev–Robatto (Chebyshev–Lobatto) nodes; evaluation uses uniform points sampled per second.

\begin{figure*}[t!]
  \centering
  \begin{minipage}[b]{0.48\textwidth}
    \centering
    \includegraphics[width=\textwidth]{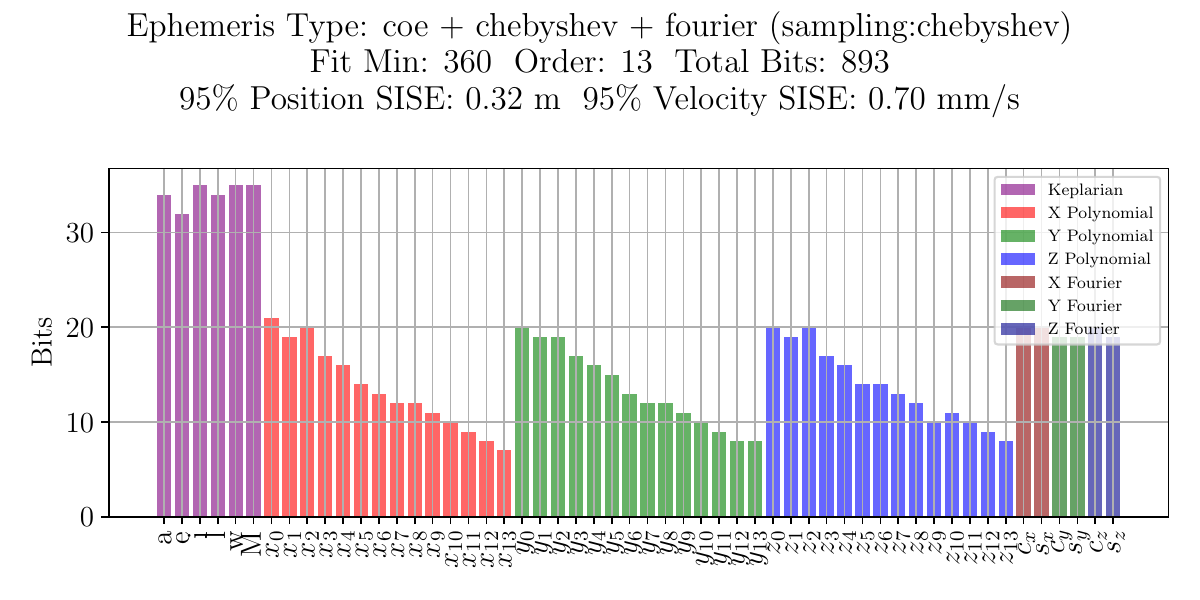}
    \caption{Bit allocation breakdown for the 30-hour ELFO (LCRNS) using Chebyshev + osculating elements + Fourier at a 360~min fitting length.}
    \label{fig:ephemfit_lcrns_bits_breakdown_fourier}
  \end{minipage}
  \hfill
  \begin{minipage}[b]{0.48\textwidth}
    \centering
    \includegraphics[width=\textwidth]{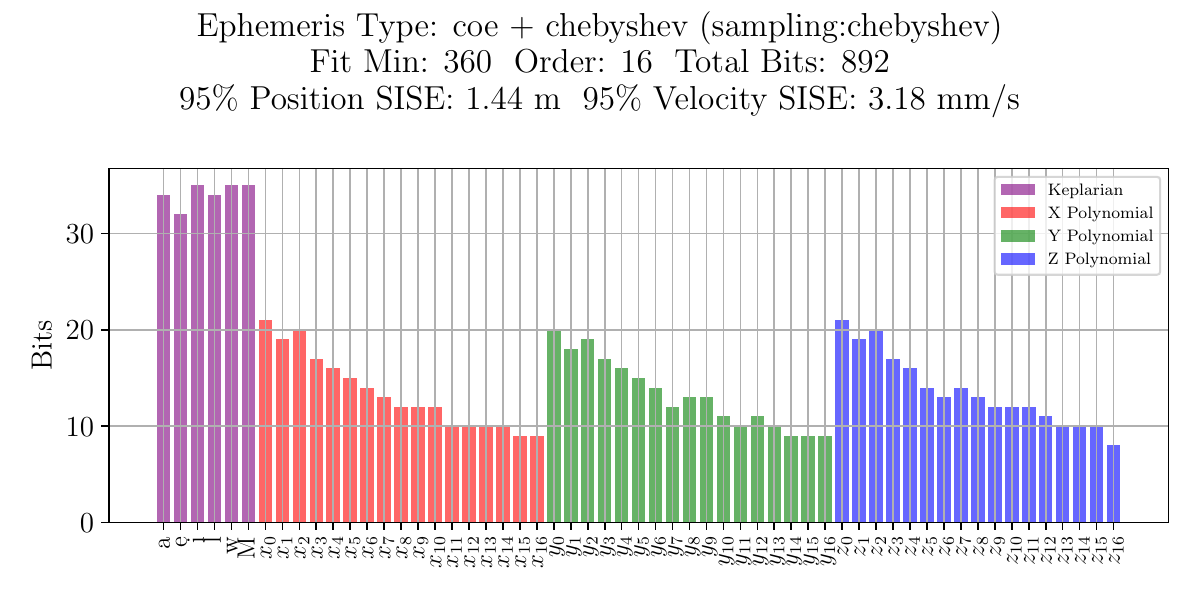}
    \caption{Bit allocation breakdown for the 30-hour ELFO (LCRNS) using Chebyshev + osculating elements at a 360~min fitting length.}
    \label{fig:ephemfit_lcrns_bits_breakdown}
  \end{minipage}
\end{figure*}

\begin{table*}[t!]
\caption{Summary of ephemeris fitting results within a 900-bit budget (95th-percentile errors).}
\label{tab:ephemeris_fitting_summary}
\centering
\resizebox{\textwidth}{!}{%
\begin{tabular}{l|cccc|cccc|cccc}
\hline
\textbf{Orbit} & 
\multicolumn{4}{c|}{\textbf{120 min}} &
\multicolumn{4}{c|}{\textbf{240 min}} &
\multicolumn{4}{c}{\textbf{360 min}} \\ 
\cline{2-13}
 & \textbf{Position (m)} & \textbf{Velocity (mm/s)} & \textbf{Bits} & \textbf{Order} 
 & \textbf{Position (m)} & \textbf{Velocity (mm/s)} & \textbf{Bits} & \textbf{Order} 
 & \textbf{Position (m)} & \textbf{Velocity (mm/s)} & \textbf{Bits} & \textbf{Order} \\
\hline
\newtext{30 hr ELFO} & 
1.12E$-$04 & 1.86E$-$02 & 654 & 19 &
5.02E$-$03 & 2.68E$-$02 & 893 & 18 &
3.17E$-$01 & 6.98E$-$01 & 893 & 13 \\
\newtext{24 hr ELFO} & 
5.37E$-$04 & 2.80E$-$02 & 816 & 19 &
5.53E$-$02 & 2.12E$-$01 & 893 & 17 &
7.63E$-$01 & 1.63E$+$00 & 882 & 13 \\
\newtext{12 hr ELFO} &
1.30E$-$02 & 1.00E$-$01 & 882 & 16 &
2.44E$-$01 & 9.01E$-$01 & 893 & 15 &
2.28E$+$00 & 5.32E$+$00 & 888 & 12 \\
\newtext{6 hr Polar} &
7.41E$-$05 & 1.60E$-$02 & 579 & 17 &
8.32E$-$05 & 1.61E$-$02 & 837 & 29 &
2.06E$-$01 & 4.43E$-$01 & 889 & 17 \\
\hline
\end{tabular}%
}
\end{table*}

\subsection{Ephemeris Fitting Results for the 30-hour ELFO (LCRNS)}

\begin{figure*}[ht!]
    \centering
    \includegraphics[width=0.85\textwidth]{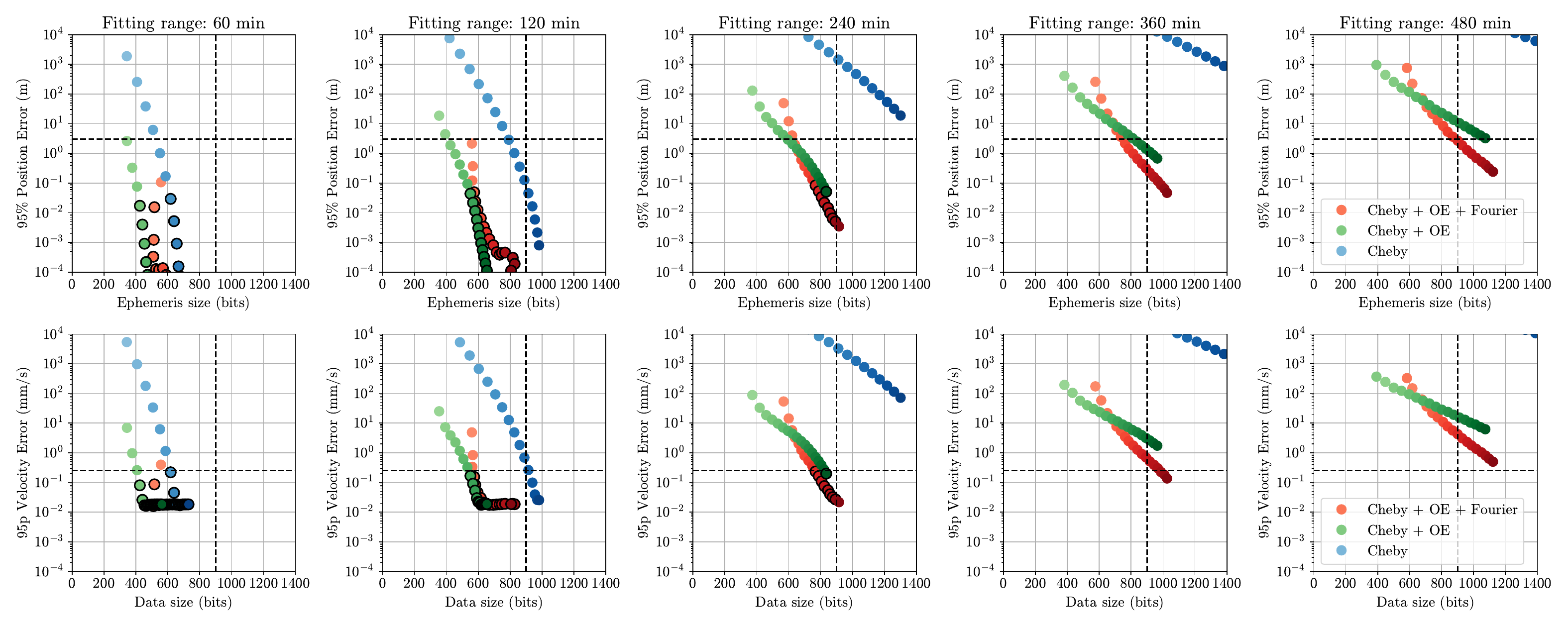}
    \newcaption{Data size versus 95th-percentile position and velocity errors for the 30-hour ELFO (LCRNS) across fitting lengths and parameterizations. 
    Blue: Chebyshev only; green: Chebyshev + osculating elements; red: Chebyshev + osculating elements + Fourier. 
    Each point corresponds to a Chebyshev order (darker = higher order). 
    Points with black outlines satisfy both error targets (3~m, 0.25~mm/s, horizontal dashed line) within 900-bits (vertical dashed line)}
    \label{fig:ephemfit_lcrns_error}
\end{figure*}

Fig.~\ref{fig:ephemfit_lcrns_error} shows the 95th-percentile position and velocity fitting errors versus data size for the 30-hour ELFO under the three parameterizations and multiple fitting lengths. 
For LCRNS we target (95th-percentile) error goals of 3~m position and 0.25~mm/s velocity, which lie comfortably within the signal-in-space error thresholds (13.43~m, 1.2~mm/s, 3\(\sigma\)) in the LunaNet SRD under a Gaussian assumption. 
The best (lowest-error) solution within 900~bits for each fitting duration and representation is summarized in Table~\ref{tab:ephemeris_fitting_summary}.

Consistent with prior work, augmenting Chebyshev polynomials with osculating elements yields markedly better accuracy at a fixed bit budget than Chebyshev-only fits. 
Moreover, for fitting ranges \(\ge\!240\)~min, adding a Fourier basis to Chebyshev + elements further reduces both position and velocity errors within the 900-bit constraint. The Fourier terms efficiently capture periodic components of the perturbations, allowing lower Chebyshev orders and smaller coefficient ranges, hence fewer bits for comparable or improved accuracy. 
For short arcs (120~min or less), the Chebyshev + elements model often provides the best accuracy per bit, as the additional Fourier term contributes limited incremental benefit over short durations.

Across fitting lengths, the velocity constraint (0.25~mm/s, 95th) is typically more stringent than the position constraint (3~m, 95th), especially for long arcs (\(\ge\!360\)~min). Under the 900-bit budget, meeting the velocity target is generally feasible for fitting lengths up to 240~min.

Fig.~\ref{fig:ephemfit_lcrns_bits_breakdown_fourier} and ~\ref{fig:ephemfit_lcrns_bits_breakdown} compare bit allocations at 360~min for the Chebyshev + OE + Fourier and Chebyshev + OE representations, respectively. 
While the Fourier model introduces roughly 120~bits for six Fourier coefficients, it enables a reduction in Chebyshev order and coefficient ranges, yielding a better overall accuracy–bit trade.

\subsection{Ephemeris Fitting Results for Other Orbits}
\begin{figure*}[t!]
    \centering
    \includegraphics[width=0.85\textwidth]{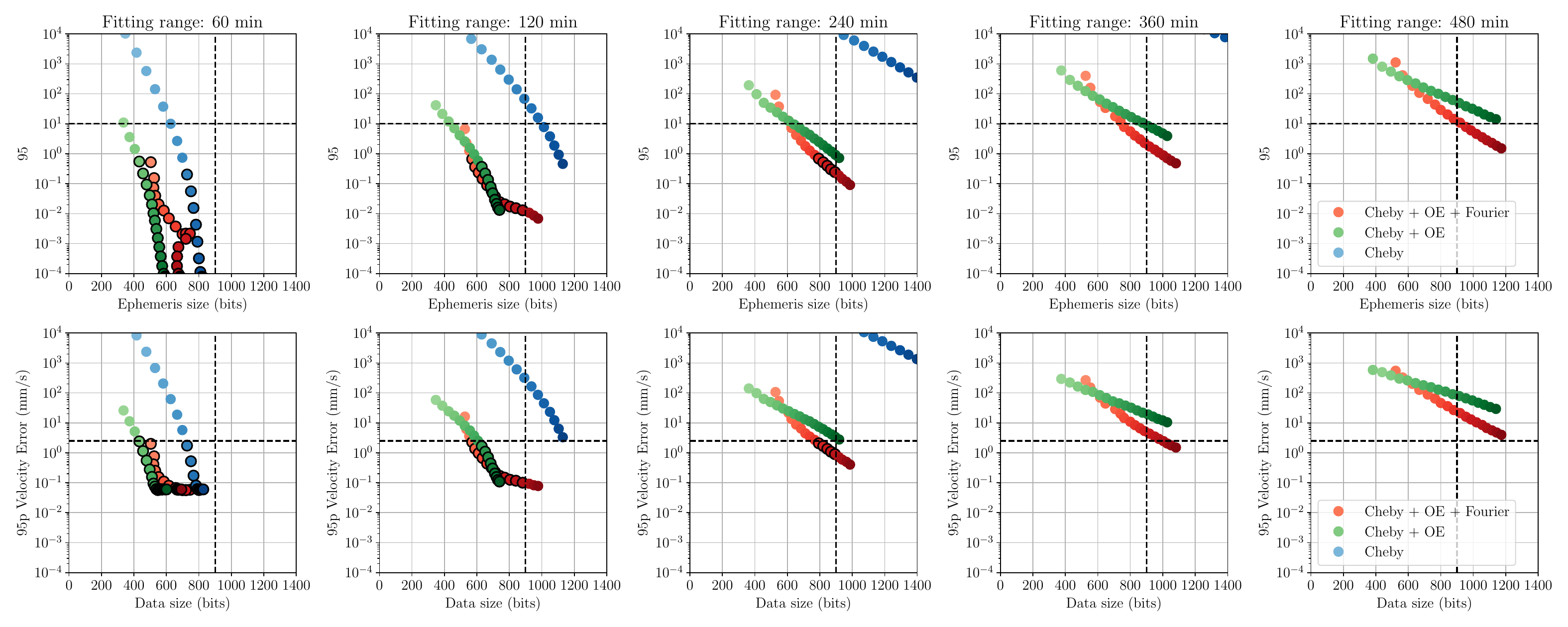}
    \newcaption{Data size versus 95th-percentile position and velocity errors for the 12-hour ELFO (LNSS) across fitting lengths and parameterizations. 
    Blue: Chebyshev only; green: Chebyshev + osculating elements; red: Chebyshev + osculating elements + Fourier. 
    Points with black outlines satisfy both error targets (10~m, 2.5~mm/s, horizontal dashed line) within 900-bits (vertical dashed line)}
    \label{fig:ephemfit_lnss_error}
\end{figure*}

\begin{figure*}[t!]
    \centering
    \includegraphics[width=0.85\textwidth]{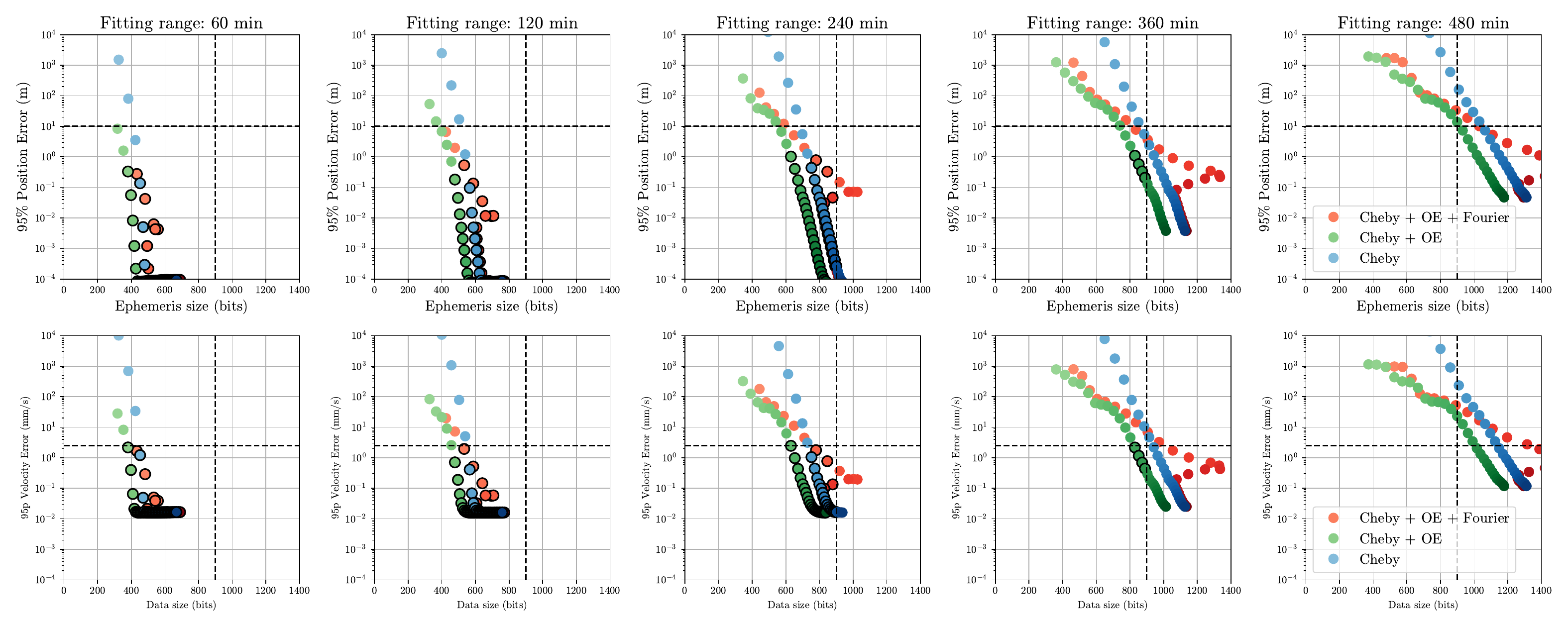}
    \newcaption{Data size versus 95th-percentile position and velocity errors for the 6-hour polar orbits. Blue: Chebyshev only; green: Chebyshev + osculating elements; red: Chebyshev + osculating elements + Fourier. 
    Points with black outlines satisfy both error targets (10~m, 2.5~mm/s, horizontal dashed line) within 900-bits (vertical dashed line)}
    \label{fig:ephemfit_polar_error}
\end{figure*}

The results for the 24-hour ELFO (LCNS), 12-hour ELFO (LNSS), and 6-hour polar orbit are summarized in Table~\ref{tab:ephemeris_fitting_summary}. 
Data size versus error plots for the 12-hour ELFO and the 6-hour polar orbit are shown in Figs.~\ref{fig:ephemfit_lnss_error} and~\ref{fig:ephemfit_polar_error}, respectively.

For ELFO regimes, trends mirror the LCRNS case: for \(\ge\!240\)~min, Chebyshev + elements + Fourier typically achieves the best accuracy within 900~bits; for shorter arcs, Chebyshev + elements is most efficient. 
For the polar orbit, Chebyshev + elements consistently outperforms the other representations across all fitting lengths, likely because short-period variations dominate and are well captured without additional Fourier terms.

As the ELFO period decreases (higher orbital speed, stronger short-period content), accuracy degrades for a fixed bit budget and fitting length, reflecting increased high-frequency variation in states and elements. 
Nonetheless, for all ELFOs we achieve sub-meter position and sub-mm/s velocity (95th) within 900~bits for fitting lengths up to 240~min; for the polar orbit, sub-meter and sub-mm/s accuracy remains achievable out to 360~min within the same bit budget.

\section{Almanac Fitting Results}
\label{sec:results_almanac}

We next evaluate the almanac model for three representative lunar orbit regimes. 
The primary role of the almanac is to support cold/warm starts by providing coarse satellite geometry sufficient to form a visibility list and predict approximate Doppler shifts~\cite{GPS_book}. 
Accordingly, we adopt a 15-day fitting interval to enable warm starts after a full lunar night (approximately 14~days). 


\subsection{Almanac Fitting Results for the 30-hour ELFO (LCRNS)}
\begin{figure*}[t!]
    \centering
    \begin{tabular}{cccc}
        \includegraphics[width=0.2\textwidth]{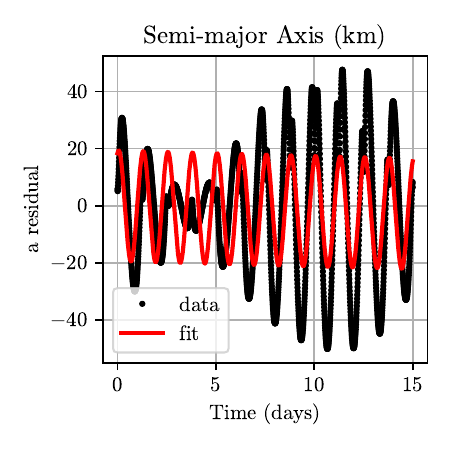} &
        \includegraphics[width=0.2\textwidth]{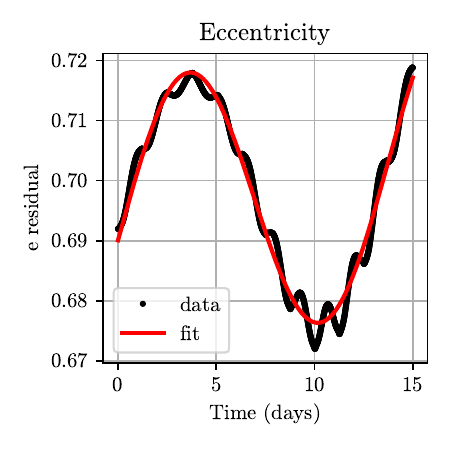} &
        \includegraphics[width=0.2\textwidth]{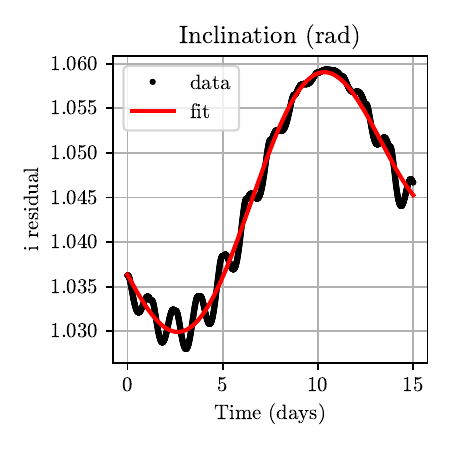} \\
        \includegraphics[width=0.2\textwidth]{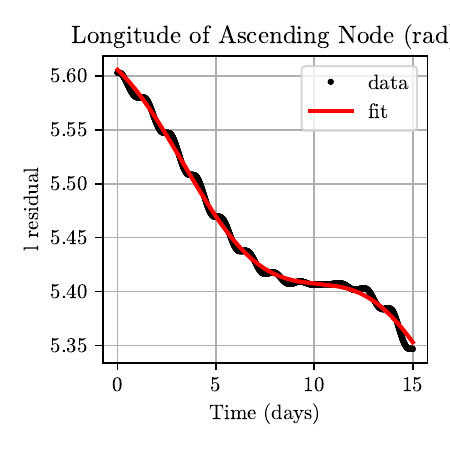} &
        \includegraphics[width=0.2\textwidth]{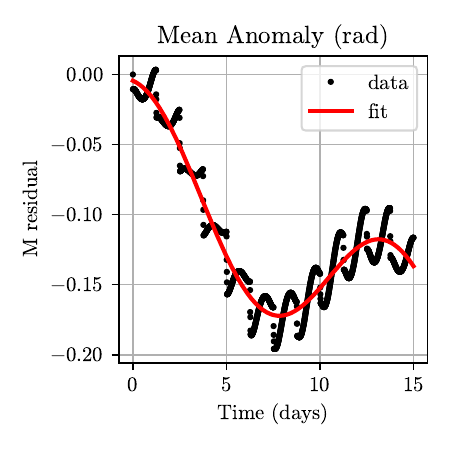} &
        \includegraphics[width=0.2\textwidth]{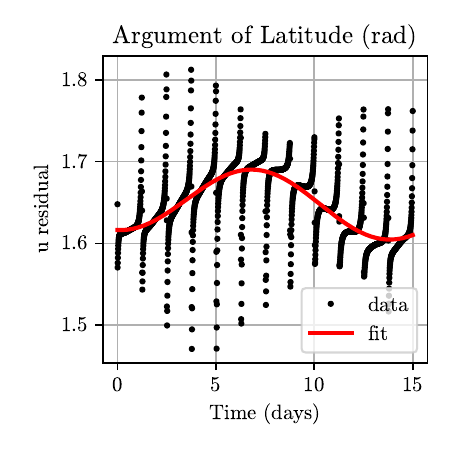}
    \end{tabular}
    \caption{\newtext{Almanac fits for the 30-hour ELFO (LCRNS) over a 15-day interval.}}
    \label{fig:almfit_lcrns}
\end{figure*}

Fig.~\ref{fig:almfit_lcrns} shows the least-squares almanac fits over 15~days for the residuals of the osculating elements \newtext{\(\{\Delta a,\Delta e,\Delta i,\Delta\Omega,\Delta M,\Delta u\}\)}. 
The model captures the dominant secular and periodic trends, but does not fully track the highest-frequency content in the semi-major axis, which manifests as along-track (tangential) phase error—visible in the mean anomaly and argument-of-latitude traces.

\begin{figure*}[ht!]
    \centering
    \includegraphics[width=0.8\textwidth]{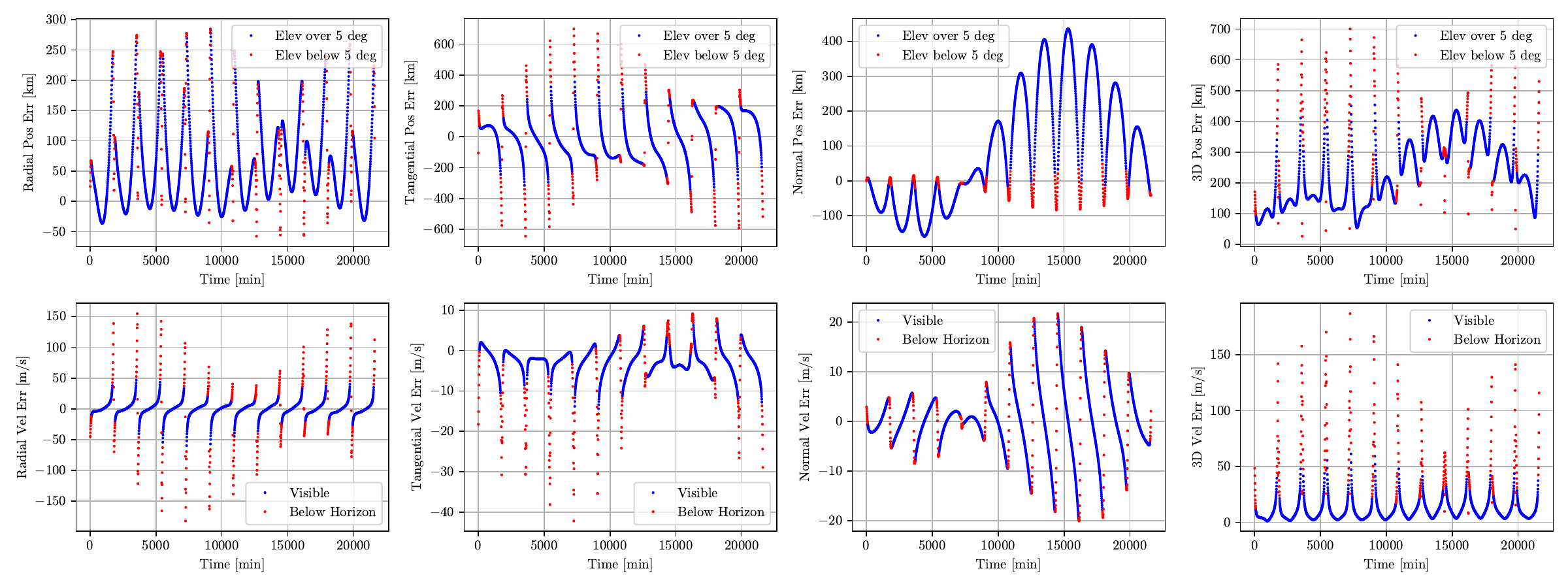}
    \newcaption{Position and velocity errors for the 30-hour ELFO (LCRNS) almanac over 15 days. Blue: epochs visible from the lunar south pole (elevation \(\geq 5^{\circ}\)); red: not visible.}
    \label{fig:almfit_lcrns_error}
\end{figure*}

Fig.~\ref{fig:almfit_lcrns_error} plots the resulting position and velocity errors over the 15-day interval. 
Error spikes occur near perilune, particularly in the along-track component, yet the overall errors remain within \(\sim 500\)~km and \(50\)~m/s across the interval when the satellite is visible from the lunar south pole.

\begin{figure*}[ht!]
    \centering
    \includegraphics[width=0.75\textwidth]{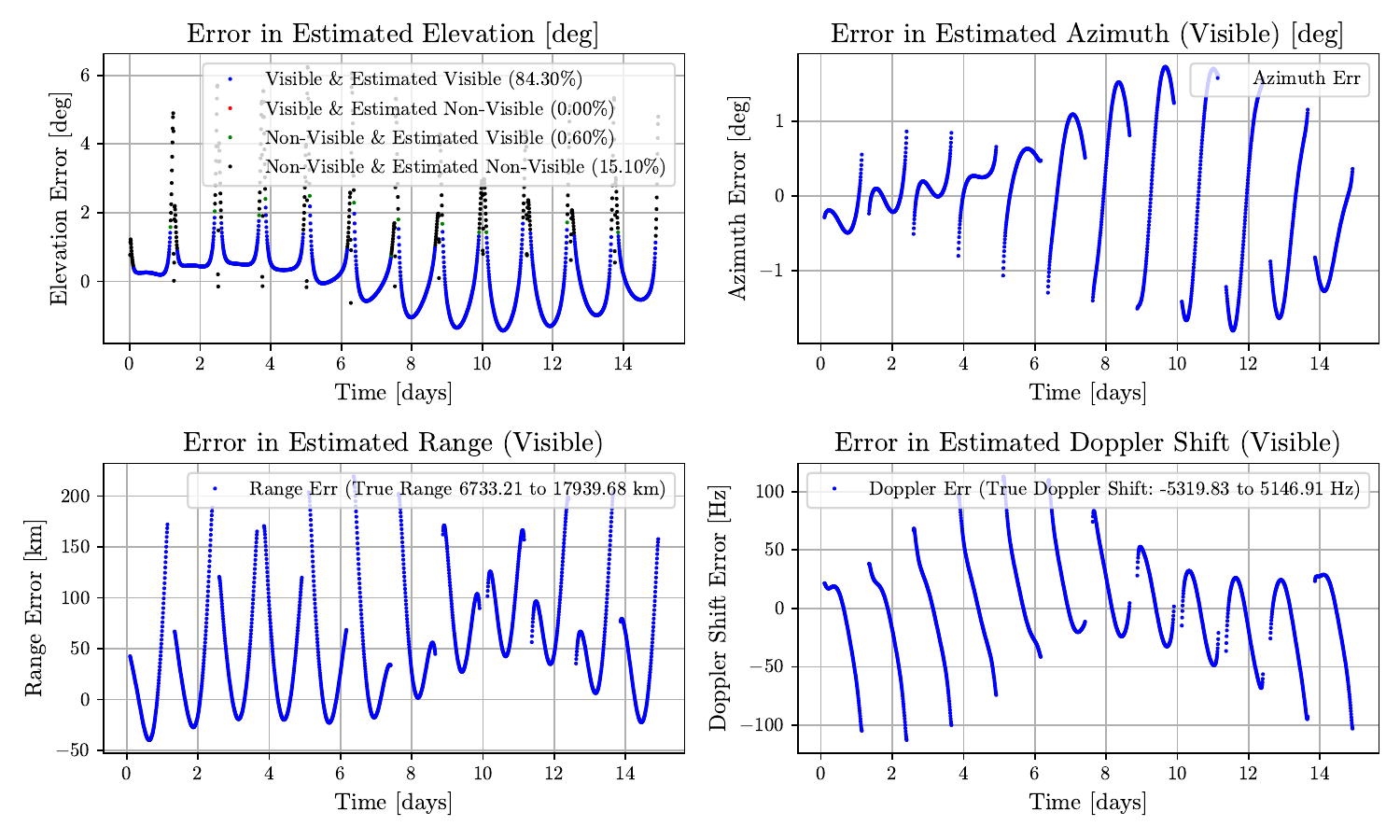}
    \newcaption{Latitude, longitude, range, and Doppler shift estimation errors with respect to the South Pole user using the fitted almanac for 30-hour ELFO (LCRNS). Only visible points from the South Pole are plotted for longitude, range, and Doppler shift.}
    \label{fig:almfit_lcrns_latlon}
\end{figure*}

Fig.~\ref{fig:almfit_lcrns_latlon} summarizes the \newtext{estimated elevation, azimuth, range, and Doppler shift} of the fitted satellite orbits. 
\newtext{
The predicted elevation is within 6 degrees of error, and can predict the visibility (elevation $\geq 5$ deg) of the satellite from the South Pole user correctly for 99.4 \% of the timesteps (except for 0.6\% of the time where the satellites are predicted visible although they are invisible).
The predicted azimuth error with respect to the South Pole is within $\pm 2$ degrees for the timesteps when the satellites are visible.
Together, the almanac provides good accuracy to judge visibility and approximately locate the satellite direction.
The range and Doppler shift estimation error with respect to the South Pole is within 200 km and 104.7 Hz in the 99th percentile, compared to the true range and Doppler shifts that take values between 6733 to 17940 km and -5320 Hz to 5147 Hz, respectively, providing information to limit the search space for acquisition.
}

\subsection{Almanac Fitting Results for Other Lunar Orbits}
Table~\ref{tab:almanac_fitting_summary} summarizes the almanac fitting results for all four cases.

\begin{table*}[htb!]
\newcaption{Summary of almanac error for 15 days. For azimuth, range, and Doppler shift, the errors are the values when satellites are visible from the South Pole user.}
\label{tab:almanac_fitting_summary}
\centering
\resizebox{\textwidth}{!}{%
\begin{tabular}{l|ccc|ccc|ccc|ccc}
\hline
\textbf{Orbit} &
\multicolumn{3}{c|}{\textbf{Elevation Error [deg]}} &
\multicolumn{3}{c|}{\textbf{Azimuth Error (Visible) [deg]}} &
\multicolumn{3}{c|}{\textbf{Range Error (Visible) [km]}} &
\multicolumn{3}{c}{\textbf{Doppler Error (Visible) [Hz]}} \\ 
\cline{2-13}
 & \textbf{Median} & \textbf{95\%} & \textbf{99\%}
 & \textbf{Median} & \textbf{95\%} & \textbf{99\%}
 & \textbf{Median} & \textbf{95\%} & \textbf{99\%}
 & \textbf{Median} & \textbf{95\%} & \textbf{99\%} \\
\hline
30 hr ELFO &
0.55 & 3.03 & 5.07 &
0.54 & 1.64 & 1.73 &
42.4 & 165.6 & 199.2 &
25.2 & 83.9 & 104.7 \\
24 hr ELFO  &
0.70 & 3.90 & 5.98 &
0.85 & 1.54 & 1.88 &
59.8 & 187.9 & 243.2 &
46.3 & 136.7 & 174.8 \\
12 hr ELFO  &
1.29 & 10.85 & 12.65 &
1.38 & 2.28 & 2.54 &
65.7 & 197.4 & 219.2 &
141.0 & 322.5 & 357.4 \\
6 hr Polar &
0.81 & 2.64 & 3.64 &
0.20 & 2.81 & 16.84 &
14.9 & 60.2 & 79.2 &
0.56 & 1.78 & 2.24 \\
\hline
\end{tabular}%
}
\end{table*}

\newtext{
For ELFOs, the angle, range, and Doppler shift estimation errors increase as the orbital period decreases (i.e., with smaller semi-major axis), with LNSS exhibiting the largest errors. Higher orbital velocities and stronger short-period perturbations amplify high-frequency variations in the orbital elements, which cannot be fully captured by a low-order almanac representation. 
}

\newtext{
Nevertheless, the 12-hour ELFO is able to predict satellite visibility from the lunar south pole for \(97.6\%\) of the time. In addition, it effectively constrains the acquisition search space by limiting the range and Doppler shift estimation errors to \(243.2\)~km and \(174.8\)~Hz at the 99th percentile, respectively. These errors are small relative to the true ranges and Doppler shifts, which span approximately \(5182\)–\(8979\)~km and \(-4918\)~Hz to \(4848\)~Hz.
}

\begin{figure*}[ht!]
    \centering
    \includegraphics[width=0.7\textwidth]{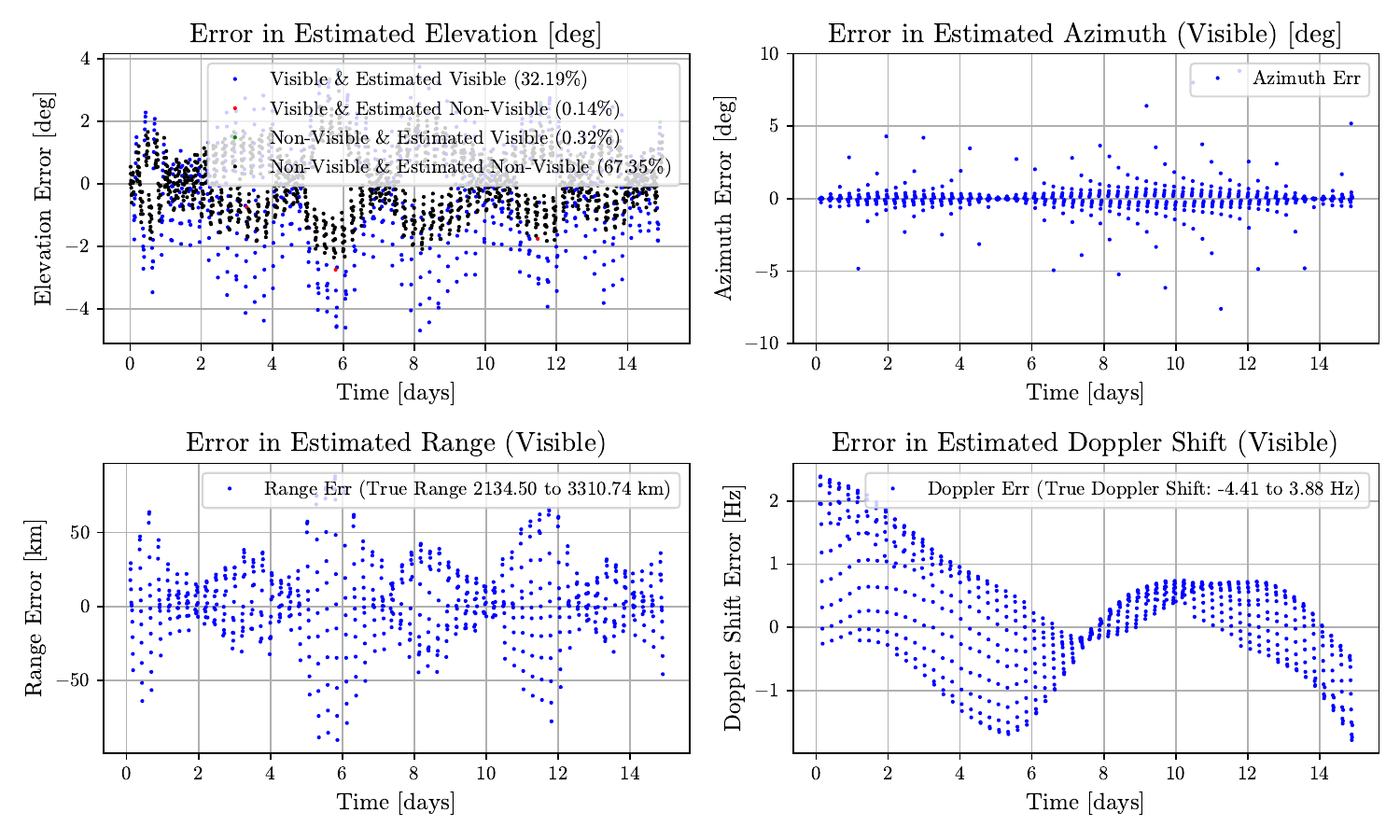}
    \newcaption{Latitude, longitude, range, and Doppler shift estimation errors with respect to the South Pole user using the fitted almanac for 6-hour polar orbit. Only visible points from the South Pole are plotted for longitude, range, and Doppler shift.}
    \label{fig:almfit_polar}
\end{figure*}

\newtext{
The errors for the 6-hour polar orbit are shown in Fig.~\ref{fig:almfit_polar}. In this case, both range and Doppler shift errors are significantly smaller than those observed for the ELFOs. This improvement is attributed to reduced perturbations from third-body gravity and lunar gravity anomalies, resulting from the smaller semi-major axis and the near-polar (\(90^\circ\)) inclination of the orbit.
}

\section{Conclusion}
\label{sec:conclusion}

This paper presented a unified framework for the design of ephemeris and almanac messages for lunar navigation satellites, addressing the unique dynamical and operational challenges of the lunar environment. By combining osculating orbital elements with Chebyshev and Fourier representations, the proposed ephemeris model achieves high-fidelity orbit reconstruction over extended fitting intervals while satisfying the LunaNet message size constraints. Across multiple orbit regimes—including 30-, 24-, and 12-hour ELFOs, as well as a 6-hour polar orbit—the hybrid representation achieves sub-meter position accuracy and sub-millimeter-per-second velocity accuracy within a 900-bit ephemeris budget for fitting intervals of up to four hours.

\newtext{For the almanac, a compact polynomial–Fourier representation of the orbital elements successfully captures long-term orbital variations over 15-day intervals. This approach yields median elevation and azimuth estimation errors below \(2^\circ\), as well as range and Doppler shift estimation errors below 10\% of their true values, thereby effectively constraining the acquisition search space for lunar navigation signals.}

Overall, the proposed framework provides a scalable and interoperable foundation for generating ephemeris and almanac products for LANS. Future work will extend this approach to incorporate orbit determination and prediction uncertainties, satellite clock models, and user receiver simulations in order to evaluate end-to-end navigation performance within the broader LunaNet architecture.

\section*{Acknowledgment}
AI tools including ChatGPT 5 was used to improve wording and correct grammar/style. AI tools were not used to generate ideas, results, figures, or analyses.

\bibliographystyle{IEEEtran}
\bibliography{references.bib}

\begin{IEEEbiography}[{\includegraphics[width=1in,height=1.25in,clip,keepaspectratio]{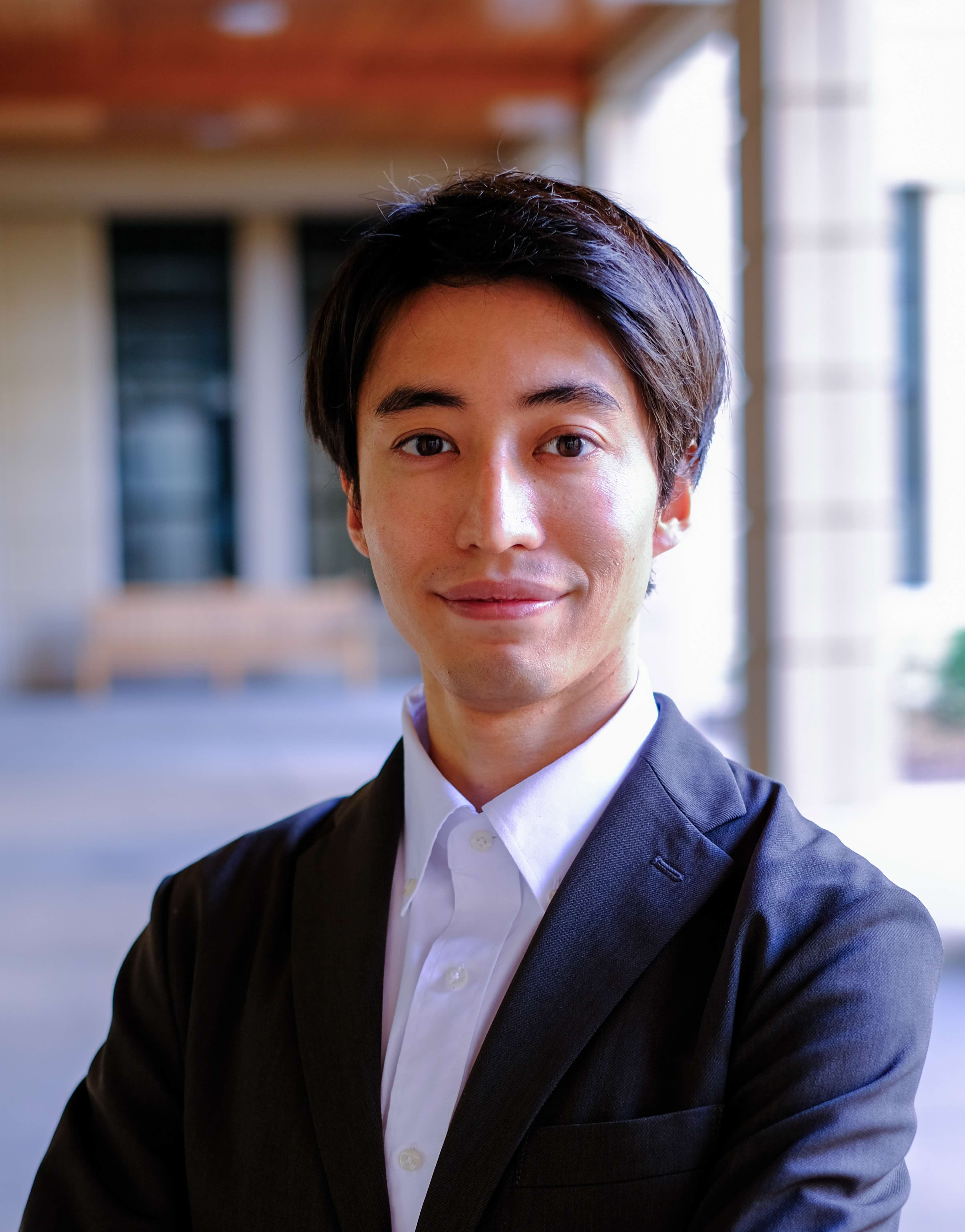}}]{Keidai Iiyama}
is a Ph.D. candidate in the Department of Aeronautics and Astronautics at Stanford University. 
He received his M.E. degree in Aerospace Engineering in 2021 from the University of Tokyo, where he also received his B.E. in 2019. 
His research focuses on positioning, navigation, and timing of lunar spacecraft and rovers, and system designs for lunar navigation systems.
\end{IEEEbiography}

\begin{IEEEbiography}[{\includegraphics[width=1in,height=1.25in,clip,keepaspectratio]{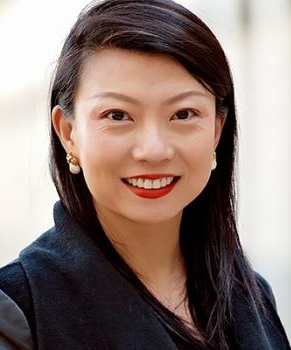}}]{Grace Gao}
is an associate professor in the Department of Aeronautics and Astronautics at Stanford University.
Before joining Stanford University, she was an assistant professor at University of Illinois at Urbana-Champaign.
She obtained her Ph.D. degree at Stanford University.
Her research is on robust and secure positioning, navigation, and timing with applications to manned and unmanned aerial vehicles, autonomous driving cars, as well as space robotics.
\end{IEEEbiography}

\end{document}